# A Perfect Terahertz Metamaterial Absorber

Alireza Bagheri Moghim
Electrical Engineering Department
Amirkabir University of Technology (Tehran Polytechnic)
Tehran, Iran
alibagheri@aut.ac.ir

Gholamreza Moradi, Associate Professor
Electrical Engineering Department
Amirkabir University of Technology (Tehran Polytechnic)
Tehran, Iran
ghmoradi@aut.ac.ir

*Abstract*—In this paper the design for an absorbing metamaterial with near unity absorbance in terahertz region is presented. The absorber's unit cell structure consists of two metamaterial resonators that couple to electric and magnetic fields separately. The structure allows us to maximize absorption by varying dielectric material and thickness and, hence the effective electrical permittivity and magnetic permeability.

## I. INTRODUCTION

Since the first theoretical [1] and experimental [2] demonstration of the properties of metamaterials (MMs), researches in this area has grown rapidly. Metamaterials have produced strange effects like negative index of refraction [3,4], and devices like electromagnetic cloak [5]. By creating independent responses to electric [6] and magnetic [1] fields, such properties could be formed. MMs have been shown in every spectral range, from radio, microwave [7], mm-Wave, THz, MIR [8], NIR, to the near optical.

Considering MMs as an effective medium [9], they can be characterized by a complex electric permittivity and magnetic permeability. Most of the work in MMs has been concentrated on real part of $\varepsilon$ and $\mu$. However, the loss components of the optical constants can be exploited as well. For example, they can be used for creating a high absorber. By using electric and magnetic resonances independently, it is possible to absorb incident electric and magnetic fields. By matching $\varepsilon$ and $\mu$, reflectivity can be minimized and metamaterial would be impedance-matched to free space. Those metamaterials that are designed to be absorbers offer benefits over conventional absorbers such as further miniaturization, wider adaptability, and increased effectiveness. Intended applications for the metamaterial absorber include emitters, sensors, spatial light modulators and etc.

The first notable trait is the thickness of the EM wave absorbers. Most are required to be at least $\lambda_0/4$ and if layers are cascaded for broader band performance this significantly increases its thickness.

We present that metamaterial can be designed to create narrow-band perfect absorbers. The simulated design reaches a peak absorptivity of over 97% at 10.02 THz.

## II. THEORY

The absorptivity of a material is given by

$$A(\omega) = 1 - T(\omega) - R(\omega), \quad (1)$$

where $T(\omega)$ is transmission and $R(\omega)$ is reflection. In terms of the complex transmissivity ($t$) and reflectivity ($r$), this equation can be written as $A = 1 - |t(\omega)|^2 - |r(\omega)|^2$. The frequency-dependent transmissivity can be written in terms of complex index of refraction $n(\omega) = n_1 + in_2$ and impedance $Z(\omega) = Z_1 + iZ_2$ for a slab of length d as

$$t(w)^{-1} = \left[\sin(nkd) - \frac{i}{2}\left(Z + \frac{1}{Z}\right)\cos(nkd)\right]e^{ikd}, \quad (2)$$

where $k = \omega/c$ and $c$ is the speed of light in the vacuum [10].

Considering impedance near unity (the free space value), the reflection decreases to zero and the equation for the transmissivity can be rewritten as follows:

$$t^{-1} = \left[\sin(nkd) - i\cos(nkd)\right]e^{ikd}. \quad (3)$$

Upon substitution of the exponential forms this becomes

$$t^{-1} = e^{-(n_1-1)kd} e^{n_2 kd}. \quad (4)$$

So the transmission ($T = |t|^2$) is

$$T = e^{-2n_2 kd}. \quad (5)$$

Therefore, as $n_2$ approaches infinity (for a given $d$),

$$\lim_{n_2 \to \infty} T = 0. \quad (6)$$

Therefore the physical derivation of the above equation is that in the absence of reflectivity, the transmission of an electromagnetic wave with the wave vector of $k$ through a slab with the thickness of $d$ is related only with losses of the slab. In order to achieve a high absorber it is necessary that the impedance to be unity and $n_2$ to be large. An important conclusion that can be made is with higher loss, the thinner slab can be in propagation direction. In this manner, the quarter-wavelength requirement of traditional absorbers can be overcame.

Precise control of *n* and *Z* is necessary to realize a high absorber. Electromagnetic metamaterials are prime candidates for this task since they can be designed to couple to electric and magnetic components of light. This enables precise tuning of the complex frequency-dependent permittivity and permeability of a metamaterial slab.

### III. DESIGN AND SIMULATION

The considered structure in this paper for terahertz metamaterial absorber, is formed of two metal elements and a substrate. The structure is shown in Fig. 1. Electric coupling is created with an electric ring resonator (ERR) as shown in Fig. 1 (a) and is presented in Ref [11]. Magnetic coupling is supplied by combination of center wire in ERR with a cut wire (Fig. 1 (b)) in a parallel plane separated by a substrate (Fig. 1 (c)), as presented in Ref [12]. So magnetic response is able to be configured by changing the dimensions of cut wire and distance between cut wire and ERR. It should be noted that electric response can be manipulated by changing the dimensions of ERR. Tuning of electric and magnetic responses separately provides an advantage that $\varepsilon$ and $\mu$ can be designed for generating a metamaterial impedance-matched to free space, minimizing reflectivity.

The simulations are performed for an ideal metamaterial absorber using the numerical solver CST [13]. The unit cell shown in Fig. 1 (c) is simulated. The appropriate boundary conditions for *xz*-plane and *yz*-plane is considered, i.e. unit cell boundary condition. From the simulation, frequency dependent S-parameters are extracted and as $R(\omega)=|S_{11}|^2$ and $T(\omega)=|S_{21}|^2$, reflectance and transmission can be achieved respectively. While 100% absorbance theoretically can be achieved, it only occurs when the metamaterial is matched to free space such that the reflectance is zero. Then by adding multiple layers, transmission can be ensured to be zero.

The simulated metamaterial had the dimensions, in micrometers, of: $a_1$=4.2, $a_2$=12, W=3.9, G=0.606, t=0.6, L=1.7 and H=11.8. The considered substrate layer for this ideal absorber has $\varepsilon_r$=4.3, $\tan\delta$=0.025 and thickness of 0.72 μm in *z*-direction. The results of simulation for the perfect terahertz metamaterial absorber is shown in Fig. 2. The simulation if performed in frequency band of 8 to 13Thz. Reflectance at 8THz is about 95% and at 13THz is 98%, but there is a minimum of a value near zero (about 0.09%) at $\omega_0$=10.02THz. Transmission at $\omega_0$ yields the value of 2.4%. Thus we achieve a total absorbance $A(\omega)=1-T(\omega)-R(\omega)$ over 97.5%. Full width at half maximum (FWHM) absorbance for the simulated metamaterial absorber is 3.5% compared to $\omega_0$.

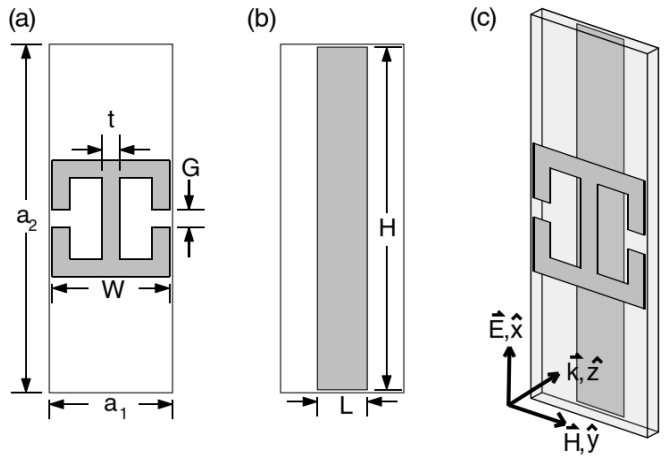

FIG. 1: Electric resonator (a) and cut wire (b). Dimension notations are listed in (a) and (b). The unit cell is shown in (c) with axes indicating the direction of propagation of a TEM wave.

We observed that the total thickness of metamaterial absorber is about 0.72 μm, while the wavelength of propagating wave at $\omega_0$ is 30 μm, it can be considered as an achievement that the thickness of absorber is less than $\lambda_0/40$. It should be noted that traditional absorbers such as Salisbury screens and Jaumann absorbers, require quarter-wavelength thickness.

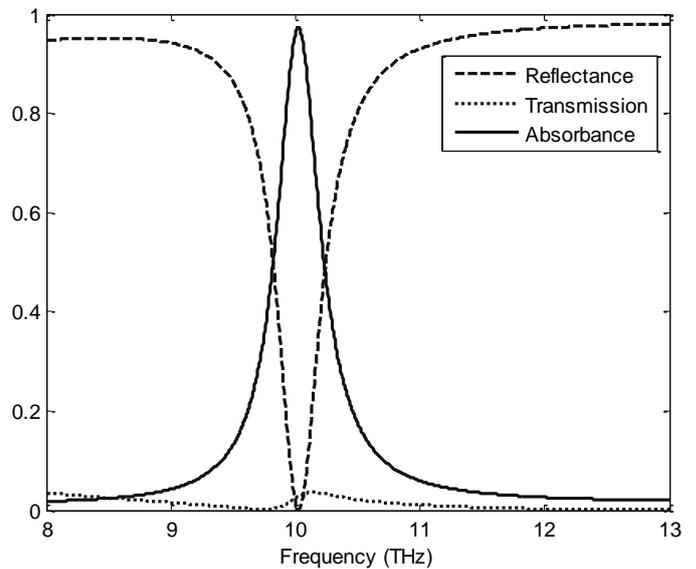

FIG. 2: Simulated perfect terahertz metamaterial absorber. Reflectance (dashed curve), transmission (pointed curve) and absorbance (solid curve) are plotted from zero to 100%.

An investigation into the effect of multiple layers of terahertz metamaterial absorbers is shown in Fig. 3. As can be observed absorbance increases sharply by adding multiple layers. Two layers of absorber achieve absorptivity of 99.74% at a frequency close to 10THz. Also for 3 layers of absorbers the absorptivity is 99.89% and for 4 layers, it is 99.91% at the peak absorbance. In the case of two layers of metamaterial absorber the total thickness in *z*-direction is 5 µm, that is still less than $\lambda_0/4$. The unit cell shown in Fig. 1 (c) can never reach complete absorptivity of 100%. It should be noted that metamaterial structure presented in this paper is not completely impedance-matched to free space.

## IV. CONCLUSION

In this paper the design and simulation for a terahertz metamaterial absorber is showed. The proposed design has absorptivity of 97.5% at 10.02THz and thickness of $\lambda_0/40$ that is a remarkable advantage for this kind of absorbers. The design presented here can be still improved. Currently, the absorber is polarization sensitive, which is not ideal for some applications.

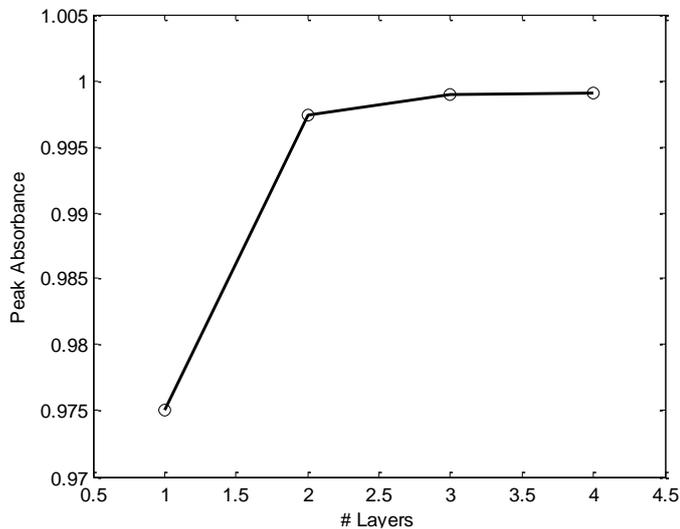

FIG. 3: Simulated absorbance with increasing metamaterial layers.